\documentclass[amssymb,prl,preprint,aps]{revtex4}
\usepackage{graphicx}
\begin{document}
\title{Phonon-assisted and magnetic field induced Kondo tunneling in
single molecular devices}
\author{K. Kikoin}
\affiliation{Scool of Physics and Astronomy, Tel-Aviv University,
Tel Aviv 69978, Israel}
\author{M.N. Kiselev}
\affiliation{The Abdus Salam International Centre for Theoretical
Physics, Trieste, Italy}
\begin{abstract}
We consider the Kondo tunneling  induced by multiphonon
emission/absorption processes in magnetic molecular complexes with
low-energy singlet-triplet spin gap and show that the number of
assisting phonons  may be changed by varying the Zeeman splitting
of excited triplet state. As a result, the structure of
multiphonon Kondo resonances may be scanned by means of magnetic
field tuning.
\end{abstract}
\maketitle

\section{Introduction}
Single electron tunneling through molecular bridges in nanodevices
is inevitably accompanied by excitation of vibrational modes.
Vibration-assisted processes usually manifest themselves in
tunneling through nanodevices as phonon satellites, which arise
around main resonance peaks (see, e.g., \cite{Flepa,KOA} and
references therein). However, phonon assistance may {\it induce}
resonance peak due to interplay with magnetic degrees of freedom
in transition metal-organic complexes (TMOC). Appearance of
Kondo-type zero bias anomaly in tunneling through TMOC was
predicted recently \cite{KKW}. In this paper we consider
multiphonon processes in Kondo tunneling in presence of magnetic
field. We discuss the fine tuning effect of magnetic field on the
Kondo tunneling induced by multiphonon processes in a situation,
where the ground state of TMOC with even occupation is a spin
singlet. In this case the Kondo resonance in tunneling arises when
the phonon emission compensates the gap between one of projections
of the excited high-spin and the ground zero spin state of TMOC.
Then the Kondo-like effect arises due to singlet/triplet
transitions, which can be treated as effective spin-flip processes
\cite{PAK}.

\section{Model}

Phonon-assisted electron tunneling through TMOC as well as
tunneling through other nanoobjects (quantum dots, nanotubes, ets)
is described within a framework of Anderson model supplemented
with the terms describing vibrational degrees of freedom and their
interaction with electron subsystem \cite{KOA,KKW,WN}
\begin{equation}\label{h1}
H = H_d + H_l + H_{tun} + H_{vib}+ H_{e-vib}~.
\end{equation}
Here $H_d$ stands for the electrons in the $3d$-shell of TMOC. It
includes strong Coulomb and exchange interaction, which
predetermine the spin quantum numbers of the corresponding
electron configuration $3d^n$ ($n$ is even)), $H_l$ contains
electrons in the metallic leads playing role of source $(s)$ and
drain $(d)$ in the electric circuit, $H_{tun}$ is responsible for
electron tunneling between TMOC and leads, $H_{vib}$ describes
vibrational degrees of freedom in TMOC, and the interaction
between electronic and vibrational subsystems is given by the last
term $H_{e-vib}$. Following the approach developed in
\cite{KOA,KKW} we choose the phonon-assisted tunneling as a source
of this interaction and represent vibrational subsystem by a
single Einstein mode $\Omega$. Then three last terms in (\ref{h1})
are written as
\begin{equation}\label{h2}
 H_{vib}+H_{tun}+ H_{e-vib}=\hbar \Omega b^\dag b +
 \sum_{ak\sigma}\left(
 te^{-\lambda(b^\dag -b)}c^\dag_{ak\sigma}d_\sigma + {\rm
 H.c.}\right).
\end{equation}
Here the operators $b$ stand for phonons, $c_{ka\sigma}, d_\sigma$
describe electrons in the leads $a=s,d$ and the d-shell of TM ion,
respectively,
$\lambda$ is the electron-phonon coupling constant. When deriving
(\ref{h2}) we assumed that the coupling is strong enough, so that
multiphonon processes are treated exactly by means of the
Lang-Firsov canonical transformation (cf. \cite{KOA}). Another
canonical transformation of Schrieffer-Wolff (SW) type \cite{Hew}
excludes $H_{tun}$ from the Hamiltonian (\ref{h1}) and maps it on
an effective Hilbert subspace with fixed (even) number of
electrons, singlet ground state and low-lying triplet excited
state of TMOC. The effective SW-like Hamiltonian is
\begin{equation}\label{h3}
{H}_{eff}=\frac{1}{2}\Delta {\bf S}^2 +hS_z +H_{l} +
   {\hat J}_T {\bf S}\cdot {\bf s}
 + {\hat J}_R {\bf R}\cdot {\bf s}
 + H_{vib}
\end{equation}
(see \cite{KKW,KA01}). Here $H_{d}$ [the first two terms in
(\ref{h3})] is represented only by spin degrees of freedom,
$\Delta=E_T-E_S$ is the energy of singlet-triplet (S-T)
transition, ${\bf S}, S_z$ is the spin operator and its
$z$-projection, $h=\mu_BgB$ is the parameter of Zeeman splitting
in external magnetic field $B$. The electron spin operator is
given by the conventional expansion $ {\bf
s}=\frac{1}{2}\sum_{kk'}\sum_{\sigma\sigma'}c^\dag_{k\sigma}{\mbox{\boldmath
$\tau$}}_{\sigma \sigma'} c_{k'\sigma'}$ where ${\mbox{\boldmath
$\tau$}}$ is the Pauli vector. Unlike the conventional SW
Hamiltonian, our Hamiltonian (\ref{h3}) contains one more vector
${\bf R}$, which describes three components of S-T transitions
\cite{KA01}. This vector will be specified below. The
electron-phonon interaction is now built into the effective
exchange constants ${\hat J}_T=t^2e^{-2\lambda(b^\dag -b)}/\delta
E_T$ and ${\hat J}_R=t^2e^{-2\lambda(b^\dag -b)}/\delta E_S$. Here
$\delta E_T,~\delta E_S$ are the energies of addition of an
electron from the leads to the TM ion in a triplet and singlet
states, respectively.

\section{Multiphonon processes in Kondo cotunneling}

The essence of the mechanism of phonon-assisted Kondo cotunneling
as it was formulated in Ref. \cite{KKW}, is that the spin
excitation energy gap $\Delta$ which quenches the conventional
Kondo effect in a nanoobject with even occupation, may be
compensated by the energy of virtual phonon emission/absorption
processes. As a result the zero bias anomaly (ZBA) arises in
tunnel conductance in spite of the zero spin ground state of the
nanoparticle. This mechanism implies fine tuning $|\hbar\Omega -
\Delta| <E_K$, where $E_K$ is the characteristic Kondo energy,
which is rather restricttive condition.
\begin{figure}[h]
\begin{minipage}{11pc}\begin{center}
\includegraphics[width=7pc]{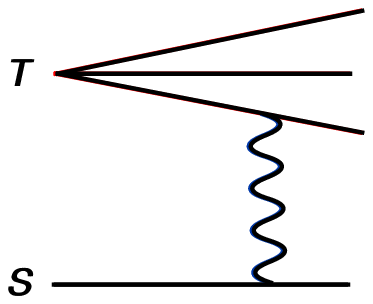}\end{center}
\caption{\label{f1} Single phonon connects singlet with spin 1
projection of triplet.}
\end{minipage}\hspace{2pc}%
\begin{minipage}{11pc}\begin{center}
\includegraphics[width=11pc]{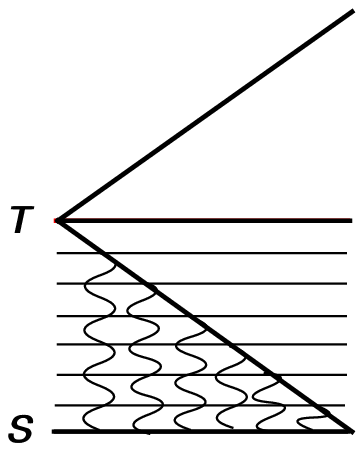}\end{center}
\caption{\label{f2}  $n$-phonon processes connect singlet with
spin 1 projection of triplet $(n\hbar\Omega<\Delta)$.}
\end{minipage}\hspace{2pc}%
\begin{minipage}{11pc}\begin{center}
\includegraphics[width=11pc]{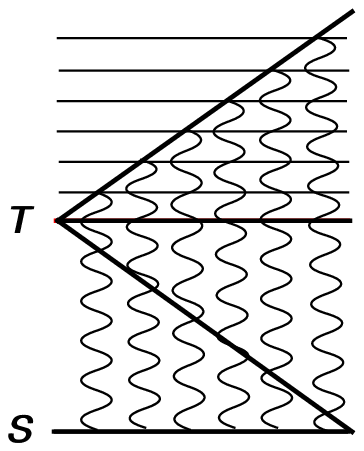}\end{center}
\caption{\label{f3} $(n+m)$-phonon processes connect singlet with
spin $\bar 1$ projection of triplet $((n+1)\hbar\Omega>\Delta)$.}
\end{minipage}
\end{figure}
To make situation more flexible, we address here to the case of
strong electron-phonon interaction and apply magnetic field as an
additional tuning instrument. The main ideas are illustrated by
Figs. 1-3. The Zeeman term in the Hamiltonian (\ref{h3}) is
responsible of splitting of the triplet state $|T_\mu\rangle$ into
3 components with spin projections $\mu=1,0,\bar1$. If the
condition $\Delta - \hbar\Omega -h  <E_K$ is satisfied (Fig.
\ref{f1}), then the states $|S\rangle, |T_1\rangle$ form effective
vector operator $\bf R$ with components
$R^+=\sqrt{2}|T_1\rangle\langle S|$, $R^-=\sqrt{2}|S\rangle
\langle T_1|$, $R_z= |T_1\rangle\langle T_1| -|S\rangle\langle
S|$, which acts effectively as a spin 1/2 operator and enters the
SW Hamiltonian (\ref{h3}) (see \cite{PAK,KA01} for further
details). The term ${\hat J}_R {\bf R}\cdot {\bf s}$ is
responsible for Kondo-type resonance tunneling in accordance with
the mechanism proposed in \cite{KKW} for zero magnetic field,
where all three components of spin $S=1$ are involved.  If the
resonance condition is fulfilled for $n$-phonon processes, $\Delta
-h - n\hbar\Omega  <E_K,$ then the Kondo tunneling is assisted by
virtual excitation of $n$-phonon "cloud" (Fig. \ref{f2}). If the
condition $\Delta +h - (n+m)\hbar\Omega  <E_K$ is valid, then the
opposite spin projection $|T_{\bar1}\rangle$ is involved in Kondo
tunneling, and the vector $\bf R$ has the components
$R^+=\sqrt{2}|S \rangle\langle T_{\bar 1}|$, $R^-=\sqrt{2}|T_{\bar
1}\rangle \langle S|$, $R_z= |S\rangle\langle S|-
|T_1\rangle\langle T_{\bar 1}|$.

\begin{figure}[h]\begin{center}
  \includegraphics[width=3.5cm,angle=0]{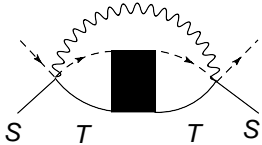}
  \caption{Phonon and parquet corrections to the vertex
    $\gamma_{B}$. Solid and dashed lines denote spin-fermion
    and conduction electron propagators,
    all parquet series are incorporated in the insertion shown by the square box.
     The multiphonon propagator is shown by the wavy line.
    }\label{fvert1}
\end{center}
\end{figure}

To find the contribution of phonon-assisted processes in Kondo
tunneling, one should calculate the exchange vertex $\gamma_h$,
which renormalizes the bare vertex $\hat J_R$ due to phonon
emission/absorption processes and logarithmically divergent parquet
insertions. This dressed vertex is shown in Fig. \ref{fvert1}. To
use the Feynman diagrammatic technique, the spin operators are
represented via effective spin-fermion operators \cite{Hew}.
The wavy line corresponds to a single-phonon
propagator in the case shown in Fig. \ref{f1}. Then the
straightforward calculation similar to that presented in
\cite{KKW} gives for the corresponding vertex
\begin{equation}\label{g1}
\gamma_h(\Omega)\sim \left[\frac{\rho
J_R^2\log\left(\displaystyle\frac{D}{{\rm
max}[k_BT,h,|\Delta-\hbar\Omega-h|]}\right)} {1-(\rho J_R
)^2\log^2\left(\displaystyle\frac{D}{{\rm
max}[k_BT,h,|\Delta-\hbar\Omega-h|]}\right)}\right]~.
\end{equation}
Here $T$ is the temperature, $D$ is the effective width of the
electron conduction band and $\rho$ is the density of states on
the Fermi level. The tunnel transparency $\cal T$ is proportional
to $\gamma^2_h$, and it is seen from (\ref{g1}), that the Kondo
peak arises as a ZBA in $\cal T$, provided
$|\Delta-\hbar\Omega-h|\sim E_K\gg B$. If the multiphonon
processes are involved in accordance with Fig. \ref{f2}, then the
wavy line in the diagram for $\gamma_h$ corresponds to the
multiphonon propagator \cite{HN} weighted with Pekarian
distribution, and the vertex function transforms into a sum of
phonon satellites
\begin{equation}
\gamma_h = \sum_n e^{-S}\frac{S^n}{n!}\gamma_h(n\Omega).
\end{equation}
\begin{figure}[h]
\includegraphics[width=7cm]{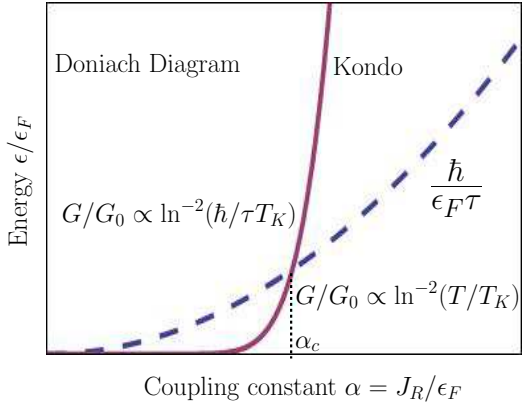}\hspace{2pc}%
\begin{minipage}[b]{19pc}\caption{Phase diagram illustrating the competition
between Kondo temperature
  and relaxation damping. Solid line stands for characteristic Kondo energy.
  Dashed line corresponds to the damping parameter.
  The phonon-assisted Kondo tunneling is effective for
  $\alpha > \alpha_c$, where $\alpha_c\sim 10^{-2}$ for TMOC coupled with metallic leads.
  $G/G_0$ is ehnancement factor for ZBA in tunnel conductance.
    }\label{fig5}
\end{minipage}
\end{figure}
Here $S=\nu/\hbar\Omega$ is the Huang-Rhys factor and
$\nu=\lambda^2/\hbar\Omega$ is the polaron shift. This equation is
valid at $kT<\hbar\Omega$ and $\nu > \rho t^2$, otherwise the
satellites smear into a single hump around the maximum of Pekarian
distribution. Since $\hbar \Omega \gg E_K,$ only one phonon
replica  satisfying the condition $|\Delta-h-n\hbar\Omega|\sim
E_K$ survive at {\it given} magnetic field $B$ . This means that
changing $B$ one may "scan" the Pekarian function in a certain
interval. In case illustrated in Fig. \ref{f3} second system of
satellites arises with Kondo peaks satisfying condition
$|\Delta+h-(n+m)\hbar\Omega|\sim E_K$.

\begin{figure}[h]\begin{center}
  \includegraphics[width=3.5cm,angle=0]{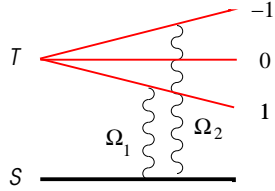}
  \caption{Two-mode regime.
    }\label{ftwo}
\end{center}
\end{figure}

Since phonon-assisted Kondo tunneling is essentially
non-equilibrium process, one should estimate the contribution of
decoherence and dephasing effects. Similar problem was discussed
in \cite{KKM,PRKW} for a situation where the gap $\Delta$ is
compensated by finite source-drain bias. To evaluate these
effects, one should calculate the damping of S-T excitation
(imaginary part of the corresponding self energy stemming from the
vertex part shown in Fig. \ref{fvert1}).  Performing calculations
in analogy with \cite{KKM}, one gets for the lifetime $\tau$ the
estimate $\hbar/\tau \sim (\rho J_R)^2 \Omega.$ This damping
should be compared with the Kondo energy extracted from
(\ref{g1}), $E_K \sim \epsilon_F\exp (-1/\rho J_R)$. The
competition between two quantities reminds that between the
indirect exchange $I_{in} \sim \rho J_R^2$ and the Kondo energy in
the Doniach diagram for Kondo lattices \cite{Hew}. However, unlike
the Doniach dichotomy, in our case $E_K$ dominates in the most
part of the phase space because $\hbar/\tau$ contains small
parameter $\Omega/D\ll 1$ comparing to $I_{in}$ (see Fig.
\ref{fig5}) Thus, the spin-phonon relaxation is not detrimental
for the phonon-assisted Kondo effect for realistic model
parameters. One more interesting situation is the case where two
vibration modes exist in a molecule, one of which may be put in a
resonance with the state $T_{1}$ at some field $h=h_1$, whereas
another one may be tuned to  the state $T_{\bar 1}$ at
$h=h_2>h_1$.   (Fig. \ref{ftwo}).

To conclude, we have demonstrated that the multiphonon
emission/absorption processes may initiate Kondo effect in TMOC
with the ground singlet state in a situation where the electron
tunneling from metallic reservoir to a $3d$ orbit of TM is
accompanied by polaronic effect in molecular vibration subsystem.
The release of vibrational energy compensates the singlet-triplet
gap in the spin excitation spectrum. Varying the Zeeman splitting
in external magnetic field, one may scan the Pekarian distribution
of Kondo-phonon satellites in tunneling conductance. Magnetic
field tuning changes drastically the character of phonon-assisted
Kondo screening. In zero magnetic field the spin $S=1$ is
underscreened by phonon-assisted Kondo processes \cite{KKW},
whereas the Zeeman splitting results in reduction of effective
spin from 1 to 1/2 and hence to complete Kondo screening. One make
hope to observe this phenomenon in the device with suspended
nanotube, which plays part of a quantum dot\cite{Hut}. The
frequency of the stretching mode in this nanotube is comparable
with the values of magnetic field used in standard experiments
($\hbar\Omega_{\rm stretch}\approx 700 \mu eV$), so one may hope
that the zero bias anomalies in Kondo tunneling assisted with one
and two phonons ($n=1$ in Fig. \ref{f3}) may be detected in this
device.
\\
Authors are greatly indebted to M R Wegewijs and A. H\"uttel for
stimulating discussions.

\end{document}